\documentclass[twocolumn,prd,showpacs,showkeys]{revtex4-1}
\usepackage{amsmath,amssymb,pgf,graphicx,color,url,array}
\usepackage{tabularx}
\usepackage{wasysym}
\usepackage{longtable}
\usepackage{ulem}
\newcommand{\be}{\begin{equation}}
\newcommand{\ee}{\end{equation}}
\newcommand{\xmax}{X_{\mathrm{max}}}
\newcommand{\pHe}{\mathrm{p}/\mathrm{He}}

\begin{document}
\begin{flushright}
\end{flushright}

\vskip -0.9cm
\title{Lower limit on the ultra-high-energy proton-to-helium ratio
  from the measurements of the tail of $X_{\mathrm{max}}$ distribution}

\author{Ivan S. Karpikov$^1$}
\author{Grigory I. Rubtsov$^1$}
\author{Yana V. Zhezher$^{1,2}$}
\email{zhezher.yana@physics.msu.ru}
\affiliation{$^1$ Institute for Nuclear Research of the Russian Academy of Sciences, 117312, Moscow, Russia,\\
$^2$ Faculty of Physics, M.V. Lomonosov Moscow State University, 119991, Moscow, Russia}

\begin{abstract}
There are multiple techniques to determine the chemical composition of
the ultra-high-energy cosmic rays. While most of the methods are
primarily sensitive to the average atomic mass, it is challenging to
discriminate between the two lightest elements: proton and helium. In
this paper, the proton-to-helium ratio in the energy range $10^{18.0}
\mbox{eV}$ to $10^{19.3} \mbox{eV}$ is estimated using the tail of the
distribution of the depth of the shower maximum
$\xmax$. Using the
exponential decay scale $\Lambda$ measured by the Pierre Auger
Observatory and the Telescope Array experiment we derive the
68\%\,CL constraints on the proton-to-helium ratio $\pHe > 7.3 $ and $\pHe >
0.43 $ for $10^{18.0} < E < 10^{18.5}$~eV and $10^{18.3} <
  E < 10^{19.3}$~eV correspondingly. It is shown that the result is conservative with
  respect to the admixture of heavier elements. We evaluate the impact
  of the hadronic interaction model uncertainty. The implications for
  the astrophysical models of the origin of cosmic rays and the safety
  of the future colliders are discussed.

\end{abstract}

\maketitle

\section{Introduction}
\label{sec:intro}

The mass composition of the ultra-high-energy cosmic rays lies among
the key tasks of major present-day and upcoming experiments. The
precise knowledge of the composition is important for understanding
the cosmic-ray production mechanism in the sources and its
population~\cite{Aloisio:2013hya}.  Moreover, composition at the
highest energies is the decisive factor for the observable flux of
cosmogenic photons~\cite{photonmodels,photonmodels2} and
neutrinos~\cite{BZ,numodels_Kotera}, see~\cite{MMWG2014} for a
review. The photons and neutrinos are more efficiently produced by the
primary protons compared to heavier elements due to the highest energy per nucleon. The diversity of the models may be illustrated with the two antipodal examples namely the dip
model~\cite{Berezinsky:2002nc,Berezinsky:2005cq,Aloisio:2006wv} and
the disappointing model~\cite{Aloisio:2009sj}. The dip model has
purely proton composition and as a consequence predicts observable
fluxes of the cosmogenic photons and neutrino. The model is named
after the dip spectral feature which is naturally explained with the
electron-positron pair production by protons. The disappointing model
includes both protons and nuclei in the source and assumes that the
acceleration of primary nuclei in the sources is rigidity-dependent
with relatively low maximum energy of acceleration. It was named
``disappointing'', because in this case there are no pion
photo-production on CMB in extragalactic space and
consequently no high-energy cosmogenic neutrino fluxes. Disappointing
model predicts no GZK-cutoff~\cite{g,zk} in the spectrum and shows no
correlation with nearby sources due to deflection of the nuclei in the
galactic magnetic fields up to the highest energies.

Another implication of the mass composition at ultra-high energies is
the investigation of safety of the future colliders. In certain
theoretical models characterized by additional spatial dimensions, the
production of non-evaporating microscopic black holes becomes possible. This
phenomenon was taken into consideration in the framework of the Large
Hadron Collider~(LHC) safety analysis~\cite{Ellis:2008hg,Giddings:2008gr}. The
proof of the LHC safety is based on the constraints on the black hole
production derived from the stability of dense astrophysical objects,
such as white dwarfs and neutron stars. The latter interact with the
ultra-high-energy cosmic rays with the center of mass energies larger
than ones achieved at LHC. One may ascertain the safety of the future
$100\ \mbox{TeV}$ colliders by studying the interaction of the cosmic
rays of the highest energies. The primary protons again play an
important role as the production of the black holes is determined by
the energy per nucleon. It was shown that the charged stable
microscopic black hole production may be excluded already, while the
exclusion of the neutral black holes would require a precise knowledge
of the proton fraction at the ultra-high
energy~\cite{Sokolov:2016lba}.

One of the most common approaches is the measurement of the
longitudinal shape of the extensive air showers (EAS). The depth of a
shower maximum, or $\xmax$, is used as a composition-sensitive
variable~\cite{Gaisser:1993ix}. The measurements of the mean $\xmax$
gives the estimate of the average atomic mass, while the study of the
$\xmax$ distribution and its moments may resolve the
multicomponent composition, see
\cite{Kampert:2012mx,composWG2012,composWG2016}.

Composition studies with the use of $\xmax$
  measurements were performed by both the Pierre Auger
  Observatory~\cite{Aab:2017njo} and the Telescope
  Array~\cite{Hanlon,Abbasi:2018nun}. Besides the derivation of the
  average atomic mass of primary particles, the data on the full shape of
  $\xmax$ distribution may be used to determine the possible
  fluxes of primary nuclei. This is performed by comparison of the
  experimental data with Monte-Carlo simulated sets.

  The Pierre Auger Observatory data set is comprised of nearly 11-year
data collected by the Fluorescence detector and the 5-year
data collected with the High Elevation Auger Telescopes (HEAT)
which extend the field of view of the Coihueco telescope
  station. The experimental data are fit jointly with the
  mixture of the proton, helium, nitrogen and iron Monte-Carlo sets
thus allowing to obtain the mass fraction of the corresponding
nuclei. The best fit imply that non-zero helium flux is
  expected in the energy range $10^{17.2} - 10^{19.5}\ \mbox{eV}$,
  while it is compatible with zero at 2$\sigma$ confidence level in
  all energy bins above $10^{18}$~eV. An improved fitting procedure
  for the Pierre Auger Data was proposed to reduce the effects of
  hadronic models uncertainties~\cite{Blaess:2018msv}. This result
  shows an indication of the presence of helium nuclei in the observed
  UHECR flux in the same energy band at somewhat higher significance.

In case of the Telescope Array, 8.5-year data from the
  Black Rock Mesa and Long Ridge fluorescence detectors operating in
  hybrid mode together with the surface detectors used. It was shown that pure protonic composition is expected in the
  energy range $10^{18.25} - 10^{19.10}\ \mbox{eV}$ at the 95\%
  confidence level. However, for higher energies the admixture of
  heavier elements can't be excluded.

The tail of the $\xmax$ distribution may be studied independently on
the main part of the distribution. It may be fit with an exponential
function $\exp (-\xmax/\Lambda)$, where $\Lambda$ is called the
attenuation length. The attenuation length is found to be sensitive to
the proton-air interaction cross-section. The first results by this
method were obtained by the Fly's Eye Collaboration~\cite{Ellsworth,
  Baltrusaitis} followed by the results of the Pierre Auger and
Telescope Array Collaborations~\cite{Auger:2012wt,Abbasi:2015fdr,Ulrich}. It
was shown in~\cite{Yushkov:2016xiz} that the attenuation length may be
used to estimate the proton-to-helium ratio $\pHe$. The latter
estimate has only minor dependence on the hadronic interaction models
and $\xmax$ experimental systematic uncertainties. 

The proton-to-helium ratio is directly measured below the knee and it
allows to constrain different astrophysical models of the origin of
cosmic rays~\cite{Biermann:1995qy, Ohira}. The measurements of the
proton-to-helium ratio at the ultra-high energies may be used
similarly to discriminate between different source models. As a recent
example, a modified dip model~\cite{Aloisio:2017kpj} confirms the
measured spectrum of the ultra-high-energy cosmic rays for the
value of proton-to-helium ratio $\pHe=5$. Furthermore,
the value of $\pHe$ used jointly with the other composition studies
will allow to pinpoint the flux of the primary protons. The latter is
an important quantity as discussed above.

The present work is dedicated to the determination of proton-to-helium
ratio of ultra-high-energy cosmic rays in the energy range from
$10^{18.0} \mbox{eV}$ to $10^{19.3} \mbox{eV}$ based on Pierre Auger Observatory and on the Telescope
Array measurements of the attenuation
length~\cite{Abbasi:2015fdr,Auger:2012wt,Ulrich}. The data are compared to the Monte-Carlo
simulations using the CORSIKA (version 7.6400) package~\cite{Heck} along with the
QGSJET II-04~\cite{Ostapchenko,QGSJETII:04} and EPOS-LHC~\cite{Pierog} hadronic
interaction models.

The paper is organized as follows. In Section~\ref{sec:method} the
analysis method is explained along with Monte-Carlo simulations. The
results on proton-to-helium ratio are presented in
Section~\ref{sec:results}. The Section~\ref{sec:conclusion} contains
concluding remarks.

\section{Method and Monte-Carlo simulations}
\label{sec:method}

The method generally follows the work of Yushkov
et. al~\cite{Yushkov:2016xiz} to derive the proton-to-helium ratio
using the measurements of the attenuation length by the Pierre Auger Observatory~\cite{Auger:2012wt} and the Telescope Array
collaboration~\cite{Abbasi:2015fdr}.

At first, the simulated sets of extensive air showers initiated by
primary protons, helium and carbon are produced with the use of the
CORSIKA package~\cite{Heck}. Simulations are performed separately with
QGSJET II-04~\cite{QGSJETII:04} and EPOS-LHC~\cite{Pierog}
hadronic interaction models for both experiments. In Auger case, for the energy range $10^{18.0} \mbox{eV} < E <10^{18.5} \mbox{eV}$ with spectral index $-3.293$~\cite{Aab:2017spctr} $17\ 098$ events are
simulated with EPOS-LHC model, and $20\ 913$ events are simulated with QGSJET II-04 model. For the Telescope Array, $17\ 354$ events are simulated for both hadronic interaction models for each species in the energy range from $10^{18.3} \mbox{eV}$ to $10^{19.3} \mbox{eV}$ with the spectrum obtained by the Telescope Array
collaboration defined by the spectral index $-3.226$ for $E <
10^{18.72}\ \mbox{eV}$ and $-2.66$ for $E >
10^{18.72}\ \mbox{eV}$~\cite{Ivanov:2015pqx}.

At the second step, the simulated sets are ``mixed'' in different
proportions from $\pHe = 0.01$ to $\pHe = 100.0$. For each mixture
the $\xmax$ distribution slope is fit exponentially to derive the
attenuation length for a mixed composition model.

An important constituent of this method is the choice of the starting
point of the fit: it can be defined in many different ways. In the
initial papers~\cite{Ellsworth, Baltrusaitis} the lower range of
$\xmax$ fit was fixed at the constant values $X_{max} = 760\ \mbox{g}
/ \mbox{cm}^2$ and $X_{max} = 830\ \mbox{g} / \mbox{cm}^2$,
respectively. Yushkov et. al~\cite{Yushkov:2016xiz} have proposed
another determination of lower fit range, which involves carbon
$\xmax$ distribution: the lower limit is defined as a value at which
only $\approx 0.5 \%$ of the carbon-initiated showers get into the
fitting range. In the present paper, the Pierre Auger instance is treated according to~\cite{Ulrich}. Experimental data analysis involves a
  three-step procedure, where first of all $\xmax$-interval containing
  $99.8\%$ of most central events is found. Then the derived
  distribution is used to obtain $\xmax$-intervals containing $20\%$
  of most deeply penetrating showers. Finally, upper end of the fit
  range is chosen to exclude $0.1\%$ of all events. This approach
  results in the following $\xmax$ fit range which is implemented
  in the presend work: $X_{\mathrm{max,start}} = 782.4\ \mbox{g} /
  \mbox{cm}^2$ to $X_{\mathrm{max,end}} = 1030.1\ \mbox{g} /
  \mbox{cm}^2$ for the $10^{18.0} - 10^{18.5} \mbox{eV}$.

For the Telescope Array case we follow the method implemented by Abbasi et al.~\cite{Abbasi:2015fdr},
where the lower limit is defined as the $X_i = \langle X_{max} \rangle
+ 40 \mbox{g} / \mbox{cm}^2$, where $\langle X_{max} \rangle$ is the
average value of a given distribution. 

Finally, after performing the fit of each mixture's $\xmax$
distribution, $\Lambda_i$ values are obtained as a function of $\pHe$
ratio. The constraints on the proton-to-helium ration are then
obtained by comparing these values with the experimental $\Lambda$
values~\cite{Abbasi:2015fdr,Auger:2012wt,Ulrich}. Lower limit on the
proton to helium ratio at $68\%\,\mbox{CL}$ corresponds to the lower
limit of the experimentally measured $\Lambda$ value is derived with the
use of measured experimental uncertainties.

\section{Results}
\label{sec:results}

\begin{figure}
\includegraphics[width=0.95\columnwidth]{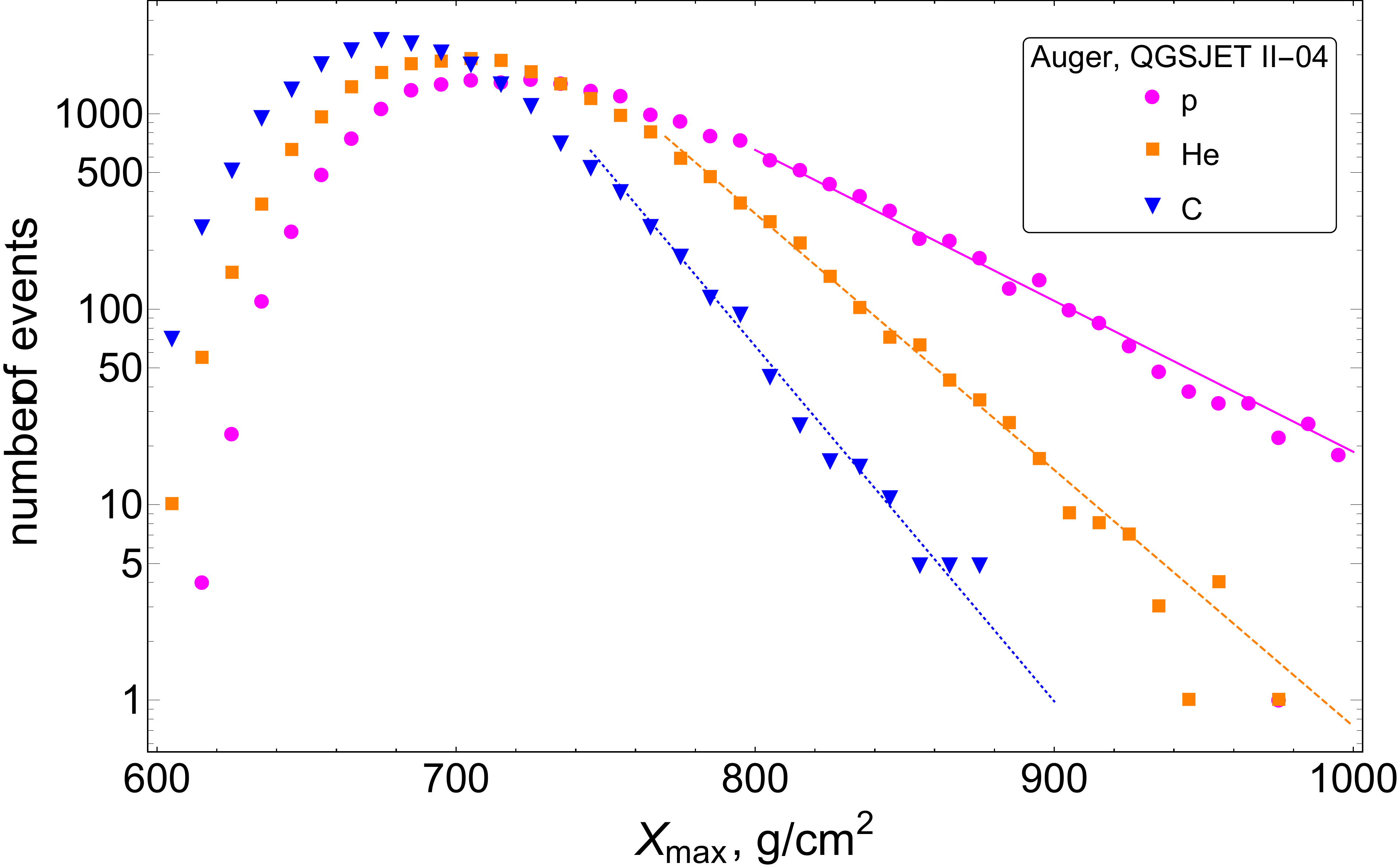}
\caption{$\xmax$ distributions for $10^{18.0} \mbox{eV} < E <
  10^{18.5}\ \mbox{eV}$ for the Pierre Auger Observatory for proton (magenta), helium (orange) and carbon
  (blue) Monte-Carlo distributions simulated with
  QGSJET~II-04. $\xmax$ distribution's tail exponential fit $\exp
  (-\xmax/\Lambda)$ is shown for each Monte-Carlo with a line of the
  corresponding color.}
\label{Xmax_PAO}
\end{figure}

\begin{figure}
\includegraphics[width=0.95\columnwidth]{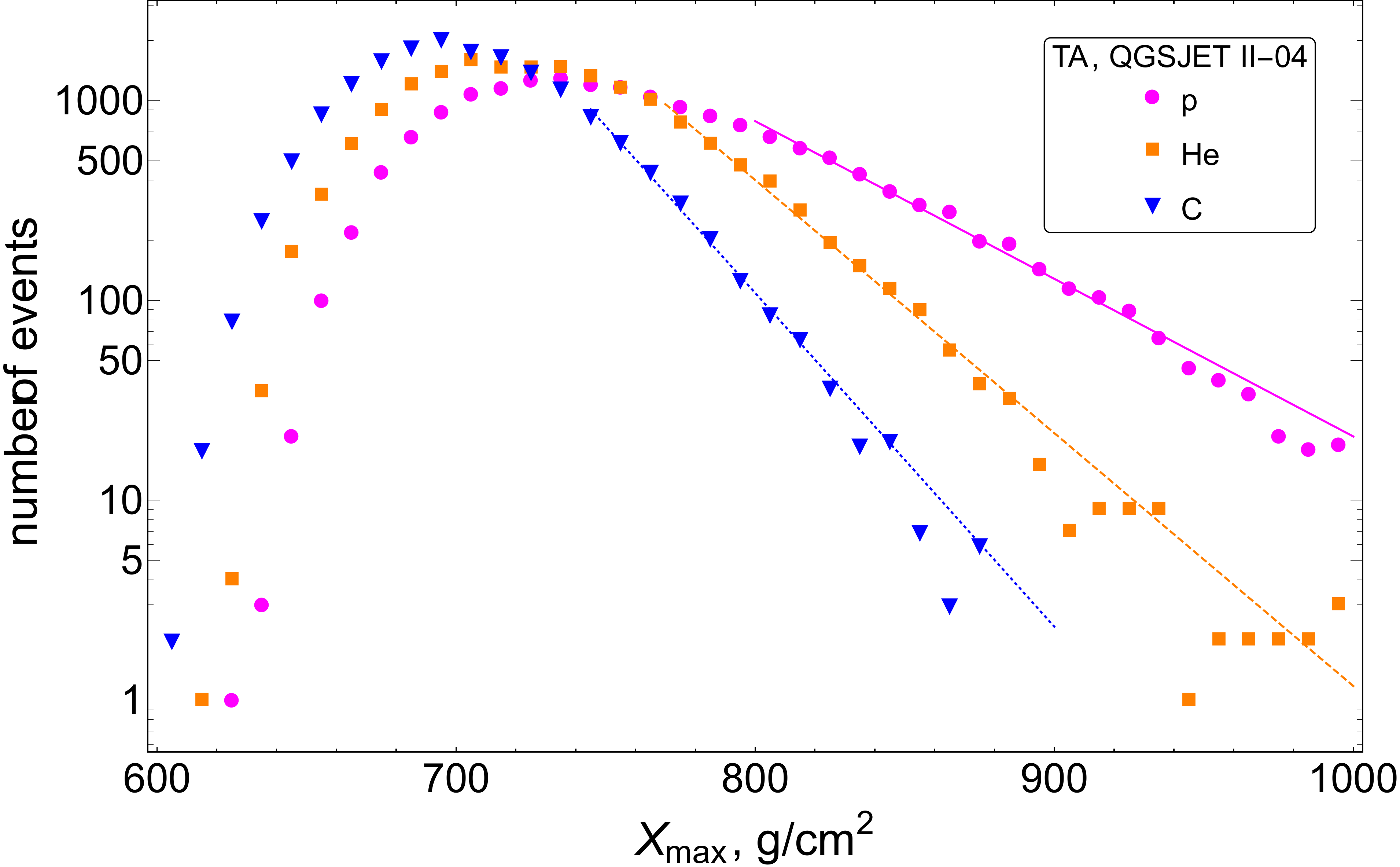}
\caption{$\xmax$ distributions for $10^{18.3} \mbox{eV} < E <
  10^{19.3}\ \mbox{eV}$ for the Telescope Array for proton (magenta), helium (orange) and carbon
  (blue) Monte-Carlo distributions simulated with
  QGSJET~II-04. $\xmax$ distribution's tail exponential fit $\exp
  (-\xmax/\Lambda)$ is shown for each Monte-Carlo with a line of the
  corresponding color.}
\label{Xmax}
\end{figure}

We present the $\xmax$ distributions and corresponding fits of exponential
tails for proton, helium and carbon Monte-Carlo simulated sets in
Figures~\ref{Xmax_PAO} and~\ref{Xmax}.

\begin{figure*}
\includegraphics[width=0.95\columnwidth]{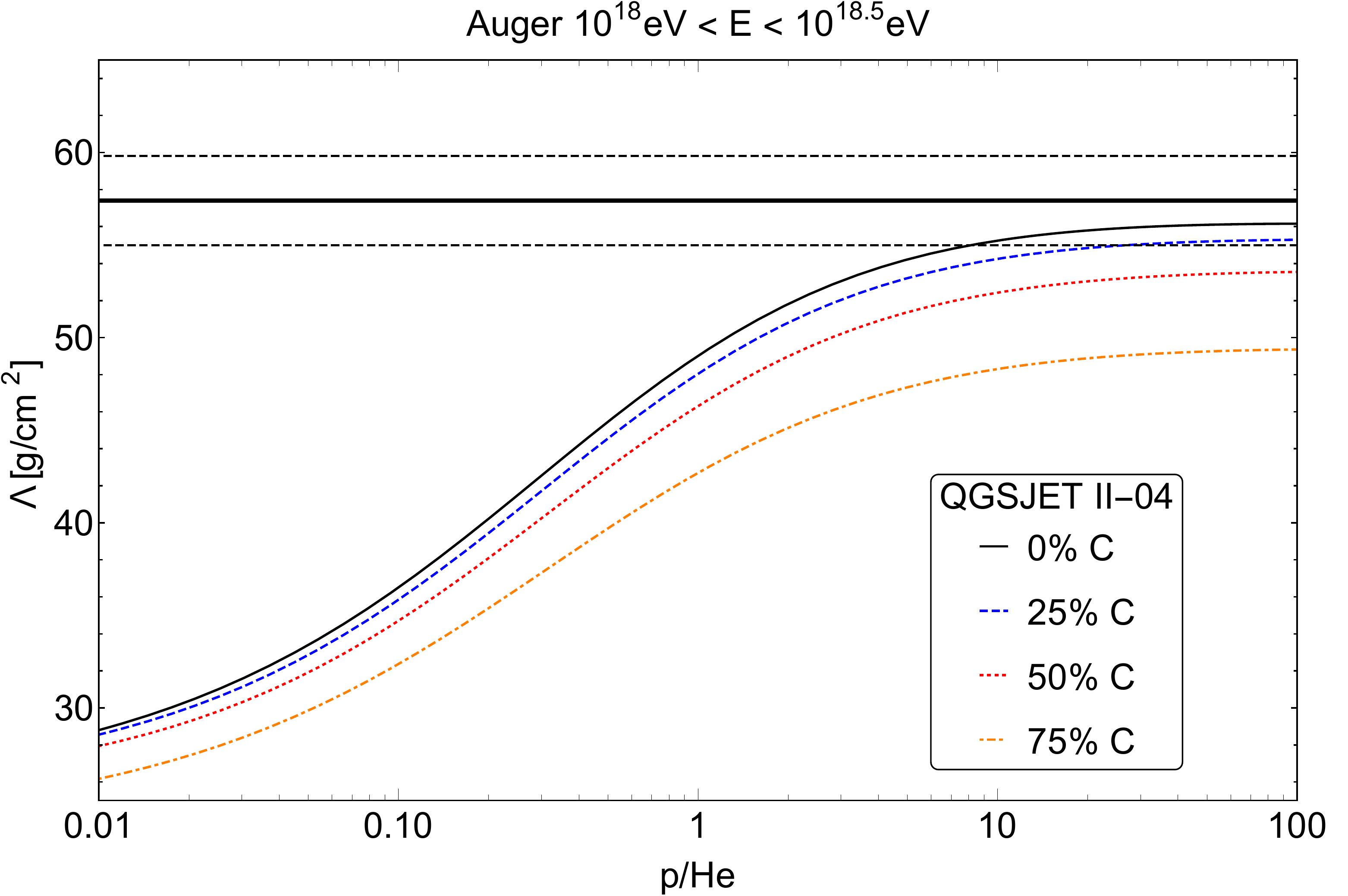}
\includegraphics[width=0.95\columnwidth]{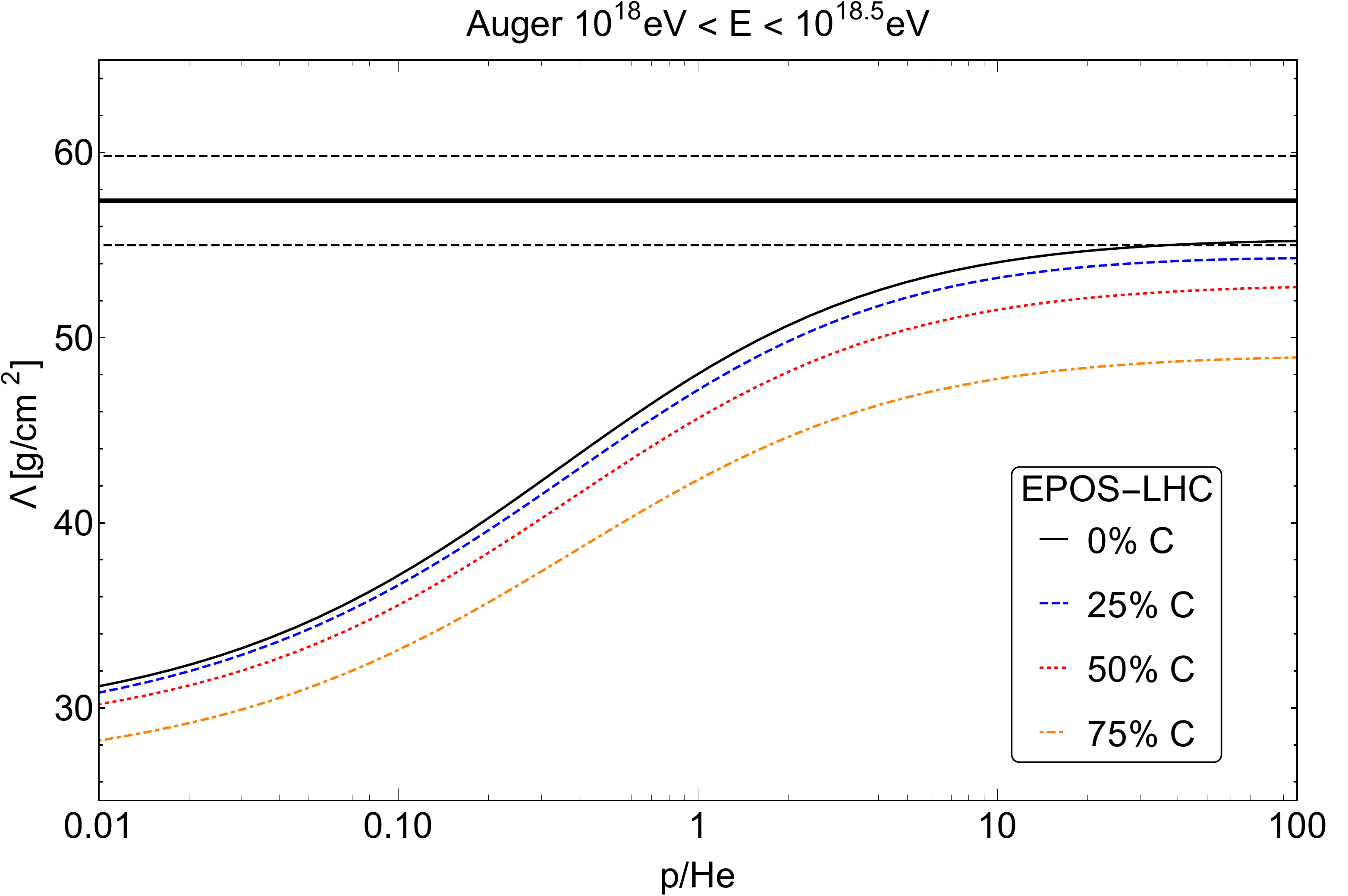}
\caption{$\Lambda$ parameter as a function of proton-to-helium ratio
  for for $10^{18.0} \mbox{eV} < E <10^{18.5} \mbox{eV}$ for two-component mixture (p and He, black line) and three
  component mixtures (p, He and 25 \% C -- green line; p, He and 50
  \% C -- red line; p, He and 75 \% C -- orange line) of Monte-Carlo events simulated with QGSJET
  II-04 (left) and EPOS-LHC (right). Black solid and dashed lines correspond to   the experimental value $\Lambda = 57.4 \pm {1.8}_{stat.} \pm {1.6}_{syst.}\ \mbox{g} /
  \mbox{cm}^2$ obtained by the Pierre Auger Observatory~\cite{Ulrich}.}
\label{pHeCPAOhigher}
\end{figure*}

\begin{figure*}
\includegraphics[width=0.95\columnwidth]{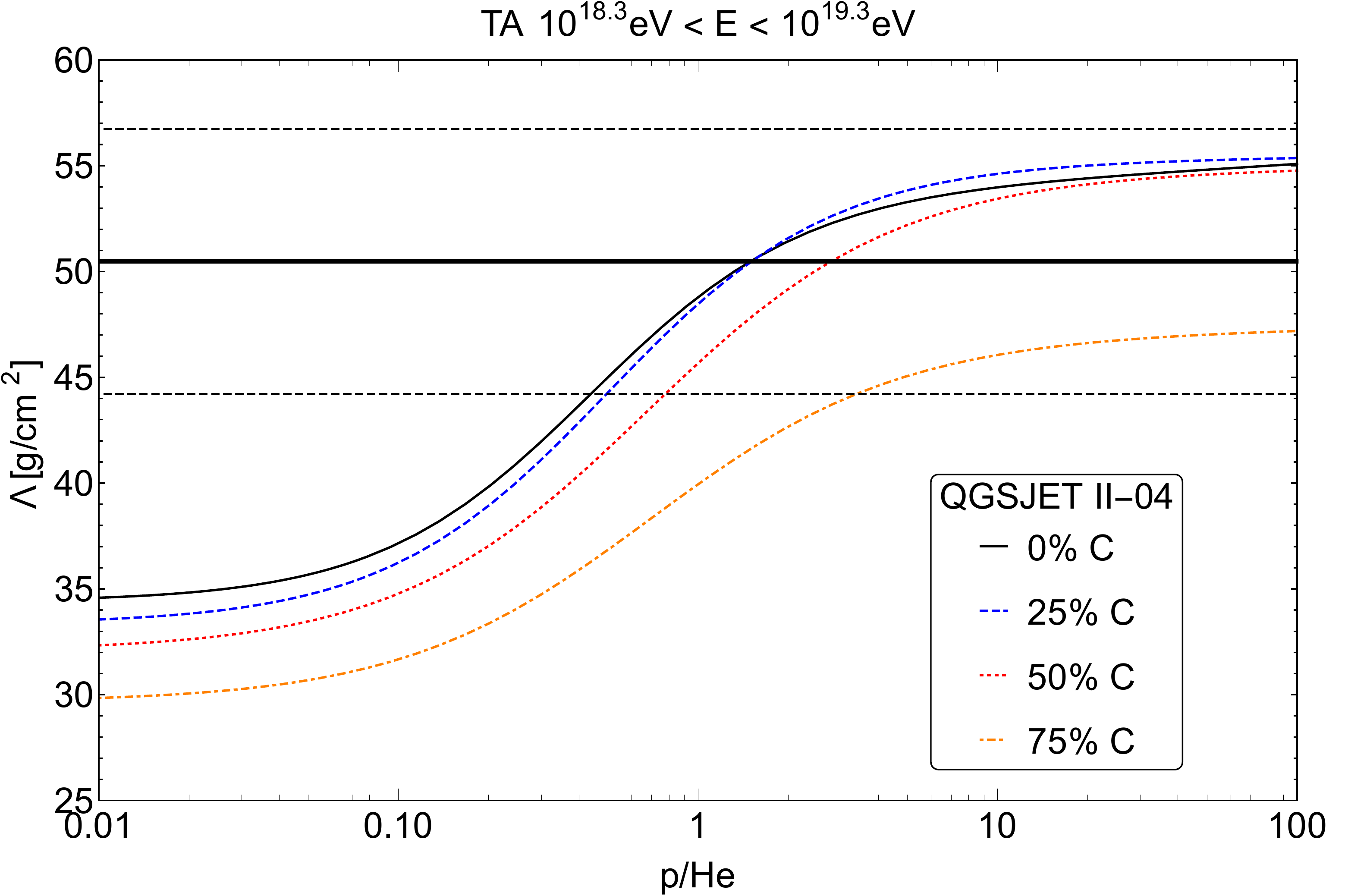}
\includegraphics[width=0.95\columnwidth]{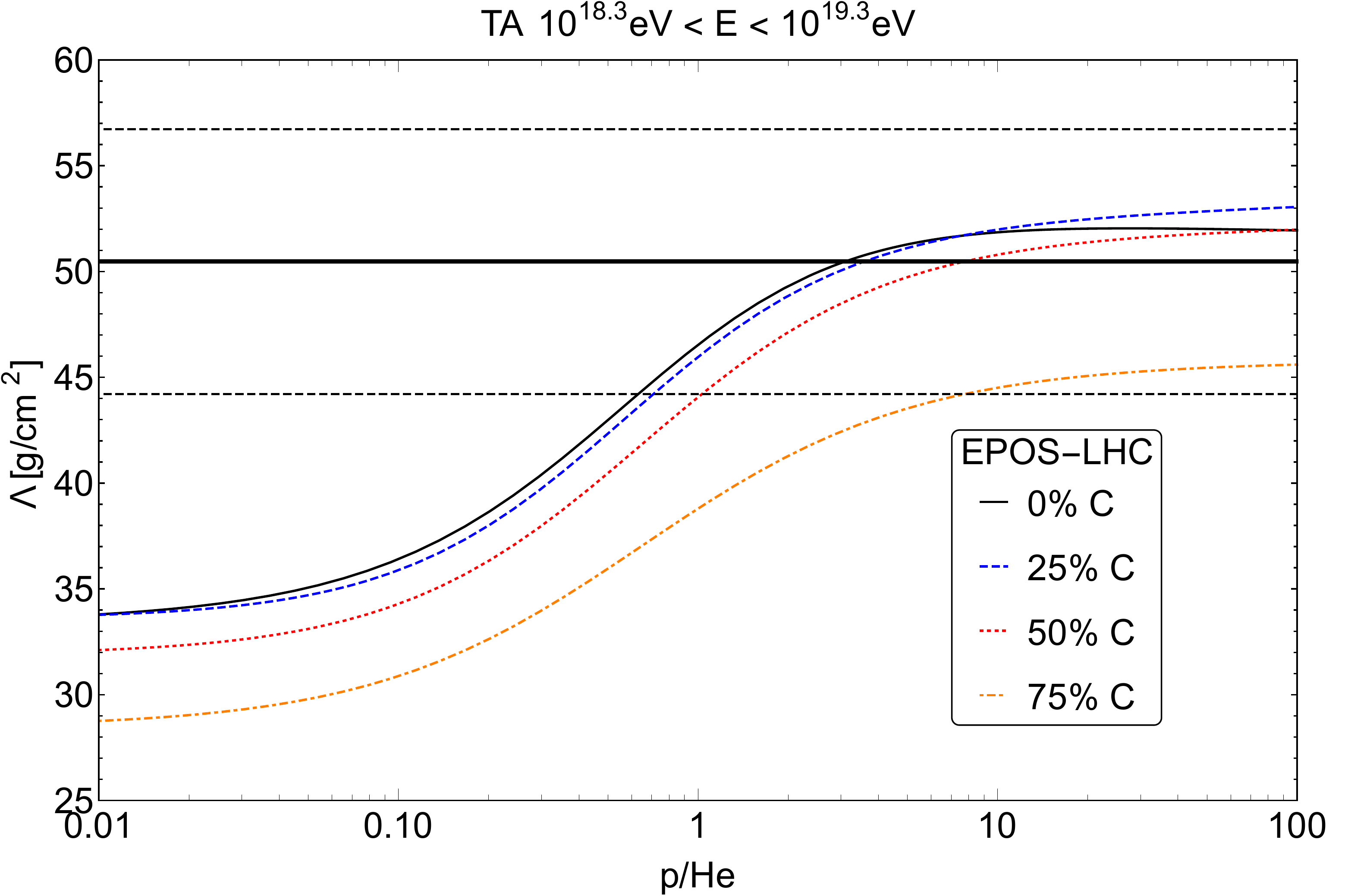}
\caption{$\Lambda$ parameter as a function of proton-to-helium ratio
  for two-component mixture (p and He, black line) and three
  component mixtures (p, He and 25 \% C -- green line; p, He and 50
  \% C -- red line; p, He and 75 \% C -- orange line) of Monte-Carlo events simulated with QGSJET
  II-04 (left) and EPOS-LHC (right). Black solid and dashed lines correspond to
  the experimental value $\Lambda = 50.47 \pm 6.26\ \mbox{g} /
  \mbox{cm}^2$ obtained by the Telescope Array
  collaboration~\cite{Abbasi:2015fdr}.}
\label{pHeC}
\end{figure*}

$\Lambda$ as a function of proton-to-helium ratio in QGSJET II-04 and
EPOS-LHC models is shown in Figures~\ref{pHeCPAOhigher} and~\ref{pHeC} with a black line. The
plot includes the proton-to-helium ratio range from $\pHe = 0.01$ to
$\pHe = 100$ with a step $\log_{10} \Delta = 0.2$. 

For the Auger case, experimental value of $\Lambda = 57.4 \pm {1.8}_{stat.} \pm {1.6}_{syst.}\ \mbox{g} /
  \mbox{cm}^2$~\cite{Ulrich} in the energy range $10^{18.0} \mbox{eV} < E <10^{18.5} \mbox{eV}$. This results in the following limits:

\begin{align}\label{limits68PAOhigher}
\pHe &> 7.3~(68\%\,\mbox{CL})~~~~~\mbox{QGSJET II-04},\\
\pHe &> 24.0~(68\%\,\mbox{CL})~~~~~\mbox{EPOS-LHC}. \nonumber
\end{align}

The TA data provides an independent measurement of the $\Lambda$ and corresponding constraints on the proton-to-helium ratio.  Comparing the Monte-Carlo function with the experimental value $\Lambda = 50.47 \pm 6.26\ \mbox{g} / \mbox{cm}^2$ obtained by the Telescope Array collaboration~\cite{Abbasi:2015fdr} in the energy range $10^{18.3} \mbox{eV} < E <
  10^{19.3}\ \mbox{eV}$ we arrive at the following lower limits on the proton-to-helium ratio:

\begin{align}\label{limits68}
\pHe &> 0.43~(68\%\,\mbox{CL})~~~~~\mbox{QGSJET II-04},\\
\pHe &> 0.63~(68\%\,\mbox{CL})~~~~~\mbox{EPOS-LHC}. \nonumber
\end{align}

We note that the pure proton composition is well compatible with the
measured attenuation length.

The stability of the method in respect to the admixture of the heavier
elements is studied. For this reason, the analysis is repeated for
three-component mixtures containing 25\%, 50\% and 75\% of carbon and
corresponding $\Lambda$ is shown in Figures~\ref{pHeCPAOhigher} and~\ref{pHeC} by the blue, red and yellow lines respectively. One may see that the constraints~\eqref{limits68PAOhigher} and~\eqref{limits68} are conservative to addition of the heavier elements as expected in~\cite{Yushkov:2016xiz}.

One may further study the three-component mixture of protons, helium
and carbon. By calculating $\Lambda$ for all possible combinations we
arrive to the following lower limits on the fraction of protons in the
three-component mixture for the Pierre Auger Observatory:

 \begin{align}\label{Climits68PAOhigher}
\mbox{p}/(\mbox{p}+\mbox{He}+\mbox{C}) &> 0.8~(68\%\,\mbox{CL})~~\mbox{QGSJET II-04},\\
\mbox{p}/(\mbox{p}+\mbox{He}+\mbox{C}) &> 0.96~(68\%\,\mbox{CL})~~\mbox{EPOS-LHC}. \nonumber
\end{align} 

\noindent And for the Telescope Array:

\begin{align}\label{Climits68}
\mbox{p}/(\mbox{p}+\mbox{He}+\mbox{C}) &> 0.20~(68\%\,\mbox{CL})~~\mbox{QGSJET II-04},\\
\mbox{p}/(\mbox{p}+\mbox{He}+\mbox{C}) &> 0.23~(68\%\,\mbox{CL})~~\mbox{EPOS-LHC}. \nonumber
\end{align}

Derived constraints are compatible with the predictions for proton flux based on $\xmax$ measurements by both the Pierre Auger Observatory and the Telescope Array~\cite{Hanlon,Abbasi:2018nun,Aab:2017njo}. Due to smaller  experimental uncertainties, the proton-to-helium ratio is derived more precisely with the Auger data.

The discussion of the possible instrumental effects is in order. The $\xmax$-distributions are known to be affected by the geometrical acceptance of the detectors as well as the reconstruction procedure, while in the scope of the \textit{Paper} $\Lambda$ values were derived for both TA and Auger assuming that protons, helium and carbon nuclei are registered with the same efficiency. In the Auger case, the unbiased $\xmax$-distribution is guaranteed by applying the fiducial cuts which extract 20\% of the most deeply penetrating showers and the events which have geometries allowing the complete observations of $\xmax$ values in the derived range~\cite{Auger:2012wt}. For the TA, it was shown in~\cite{Abbasi:2015fdr}, that there is no bias introduced to the $\Lambda$ value if either the thrown proton distribution without any detector effects is used, or the one which was propagated through the detector and then reconstructed. Moreover, as it is shown in~\cite{Abbasi:2014sfa}, the precision of the $\xmax$ reconstruction for iron Monte-Carlo events is somewhat higher than the one for proton events. No biases are expected for intermediate nuclei as well based on the common assumption that the shower properties depend smoothly on $\ln A$.

\section{Conclusion}
\label{sec:conclusion}

Let us finally discuss the possible applications of the obtained lower
limit on the proton-to-helium ratio at the energies $10^{18.0}
\mbox{eV} < E < 10^{19.3} \mbox{eV}$. First of all, we consider the
properties of the sources of the ultra-high-energy cosmic rays in the
view of the constraints~Eq.~\eqref{limits68PAOhigher} and~Eq.~\eqref{limits68}. The present limits constrain
the models with helium domination in the energy range under study,
e.g. the helium version of the disappointing
model~\cite{Aloisio:2009sj}. These models generally include the
preferential acceleration of helium or an excessive helium abundance
at the acceleration region. The result of the present {\it Paper} is fully compatible with the original pure proton dip model~\cite{Berezinsky:2002nc,Berezinsky:2005cq,Aloisio:2006wv} as long as with the standard disappointing model~\cite{Aloisio:2009sj} with $\pHe \sim 1$ while the modification of the dip model with $\pHe=5$~\cite{Aloisio:2017kpj} is disfavored by the Auger data~Eq.~\eqref{Climits68PAOhigher}.

Secondly, let us discuss the safety of the future colliders. The proof of the safety relies largely on the existence
  of non-zero flux of the ultra-high-energy protons~\cite{Sokolov:2016lba}. One may see from the Figure~\ref{pHeCPAOhigher} that expected $\Lambda$ for models with zero proton flux is more than $5$ standard deviation away from the value measured by the Pierre Auger Observatory. Hence the safe operation of the future colliders is supported at the high confidence level.

\section*{Acknowledgements}

We would like to thank L.~Bezrukov, O.~Kalashev, M.~Kuznetsov, M.~Pshirkov, A.~Sokolov and S.~Troitsky for helpful comments and suggestions. The authors are indebted to the members of the Telescope Array collaboration for inspiring discussions.

The present work is supported by the Russian Science Foundation grant
No. 17-72-20291. The numerical part of the work is performed at the
cluster of the Theoretical Division of INR RAS.



\begin{thebibliography}{99}
\bibitem{Aloisio:2013hya}
  R.~Aloisio, V.~Berezinsky and P.~Blasi,
  JCAP {\bf 1410}, no. 10, 020 (2014).
\bibitem{photonmodels}
  G.~Gelmini, O.~E.~Kalashev and D.~V.~Semikoz,
  J.\ Exp.\ Theor.\ Phys.\  {\bf 106}, 1061 (2008).
\bibitem{photonmodels2}
  D.~Hooper, A.~M.~Taylor and S.~Sarkar,
  Astropart.\ Phys.\  {\bf 34}, (2011) 340.
\bibitem{BZ}
  V.~S.~Berezinsky and G.~T.~Zatsepin,
  Phys.\ Lett.\  {\bf 28B}, 423 (1969).
\bibitem{numodels_Kotera}
  K.~Kotera, D.~Allard and A.~V.~Olinto,
  JCAP {\bf 1010}, 013 (2010).
\bibitem{MMWG2014}
  T.~Karg {\it et al.} [IceCube and Pierre Auger and Telescope Array Collaborations],
  JPS Conf.\ Proc.\  {\bf 9}, 010021 (2016).
\bibitem{Berezinsky:2002nc}
  V.~Berezinsky, A.~Z.~Gazizov and S.~I.~Grigorieva,
  Phys.\ Rev.\ D {\bf 74}, 043005 (2006).
\bibitem{Berezinsky:2005cq}
  V.~Berezinsky, A.~Z.~Gazizov and S.~I.~Grigorieva,
  Phys.\ Lett.\ B {\bf 612}, 147 (2005).
\bibitem{Aloisio:2006wv}
  R.~Aloisio, V.~Berezinsky, P.~Blasi, A.~Gazizov, S.~Grigorieva and B.~Hnatyk,
  Astropart.\ Phys.\  {\bf 27}, 76 (2007).
\bibitem{Aloisio:2009sj}
  R.~Aloisio, V.~Berezinsky and A.~Gazizov,
  Astropart.\ Phys.\  {\bf 34}, 620 (2011).
  \bibitem{g}
  K.~Greisen, Phys.\ Rev.\ Lett.\ {\bf 16}, 748  (1966)
\bibitem{zk}
  Z.~T.~Zatsepin and V.~A.~Kuz'min, Zh.\ Eksp.\ Teor.\ Fiz.\ Pis'ma Red. {\bf 4}, 144 (1966)
\bibitem{Ellis:2008hg}
  J.~R.~Ellis, G.~Giudice, M.~L.~Mangano, I.~Tkachev and U.~Wiedemann, J.\ Phys.\ G {\bf 35}, 115004 (2008), [arXiv:0806.3414 [hep-ph]].

\bibitem{Giddings:2008gr}
  S.~B.~Giddings and M.~L.~Mangano, Phys.\ Rev.\ D {\bf 78}, 035009 (2008), [arXiv:0806.3381 [hep-ph]].

\bibitem{Sokolov:2016lba}
  A.~V.~Sokolov and M.~S.~Pshirkov, Eur.\ Phys.\ J.\ C {\bf 77}, no. 12, 908 (2017), [arXiv:1611.04949 [hep-ph]].

\bibitem{Gaisser:1993ix}
  T.~K.~Gaisser {\it et al.} [HiRes Collaboration],
  Phys.\ Rev.\ D {\bf 47}, 1919 (1993).

\bibitem{composWG2012}
  E.~Barcikowski {\it et al.} [Pierre Auger and Yakutsk Collaborations],
  EPJ Web Conf.\  {\bf 53}, 01006 (2013).

\bibitem{composWG2016}
  W.~Hanlon {\it et al.},
  JPS Conf.\ Proc.\  {\bf 19}, 011013 (2018).

  \bibitem{Kampert:2012mx}
  K.~H.~Kampert and M.~Unger,
  Astropart.\ Phys.\  {\bf 35}, 660 (2012).

  \bibitem{Aab:2017njo} 
  J.~Bellido {\it et al.} [Pierre Auger Collaboration],
  PoS ICRC {\bf 2017} 506, arXiv:1708.06592 [astro-ph.HE].

    \bibitem{Hanlon}
  W.~Hanlon {\it et al.}, PoS ICRC {\bf 2017}, 536 (2017).

  \bibitem{Abbasi:2018nun} 
  R.~U.~Abbasi {\it et al.} [Telescope Array Collaboration],
  Astrophys.\ J.\  {\bf 858}, no. 2, 76 (2018), [arXiv:1801.09784 [astro-ph.HE]].

  \bibitem{Blaess:2018msv} 
  S.~Blaess, J.~A.~Bellido and B.~R.~Dawson,
  arXiv:1803.02520 [astro-ph.HE].

  \bibitem{Ellsworth}
  R.~Ellsworth {\it et al.} [Fly's Eye Collaboration], Phys.\ Rev.\ D\ {\bf 26}, 336 (1982).

  \bibitem{Baltrusaitis}
  R. Baltrusaitis {\it et al.} [Fly's Eye Collaboration], Phys.\ Rev.\ Lett.\ {\bf 52}, 1380 (1984).

  \bibitem{Abbasi:2015fdr}
  R.~U.~Abbasi {\it et al.} [Telescope Array
    Collaboration], Phys.\ Rev.\ D {\bf 92}, no. 3, 032007 (2015).

  \bibitem{Auger:2012wt}
   P.~Abreu {\it et al.} [Pierre Auger
    Collaboration], Phys.\ Rev.\ Lett.\ {\bf 109}, 062002 (2012).

            \bibitem{Ulrich}
  R.~Ulrich, PoS ICRC {\bf 2015}, 401 (2016).

          \bibitem{Yushkov:2016xiz}
  A.~Yushkov, M.~Risse, M.~Werner and J.~Krieg, Astropart.\ Phys.\  {\bf 85}, 29 (2016), [arXiv:1609.08586 [astro-ph.HE]].

  \bibitem{Biermann:1995qy}
  P.~L.~Biermann, T.~K.~Gaisser and T.~Stanev, Phys.\ Rev.\ D {\bf 51}, 3450 (1995), [astro-ph/9501001].

  \bibitem{Ohira}
  Y.~Ohira, K.~Ioka, Astrophys.\ J\ Lett.\ {\bf 729}, L13 (2011).

  \bibitem{Aloisio:2017kpj}
  R.~Aloisio and V.~Berezinsky, arXiv:1703.08671 [astro-ph.HE].

  \bibitem{Heck}
D.~Heck {\it et al.}, Report FZKA-6019 (1998), Forschungszentrum Karlsruhe.

\bibitem{Ostapchenko}
  S.~Ostapchenko, Nucl.\ Phys.\ Proc.\ Suppl.\  {\bf 151}, 143 (2006) [hep-ph/0412332].

\bibitem{QGSJETII:04}
  S.~Ostapchenko,
  Phys.\ Rev.\ D {\bf 83}, 014018 (2011)

   \bibitem{Pierog}
  T.~Pierog, I.~Karpenko, J.~M.~Katzy, E.~Yatsenko and K.~Werner, Phys.\ Rev.\ C {\bf 92}, no. 3, 034906 (2015) [arXiv:1306.0121 [hep-ph]].

  \bibitem{Aab:2017spctr} 
  F.~Fenu {\it et al.} [Pierre Auger Collaboration],
  PoS ICRC {\bf 2017} 486, arXiv:1708.06592 [astro-ph.HE].

  \bibitem{Ivanov:2015pqx}
  D.~Ivanov, PoS ICRC {\bf 2015}, 349 (2016).

  \bibitem{Abbasi:2014sfa} 
  R.~U.~Abbasi {\it et al.},
  Astropart.\ Phys.\  {\bf 64}, 49 (2015) [arXiv:1408.1726 [astro-ph.HE]].


\end{thebibliography}
\end{document}